\documentclass[12pt]{article}
\usepackage{amsfonts}
\usepackage{latexsym}
\usepackage{amsmath}
\usepackage{amssymb}
\usepackage{amssymb}
\usepackage[utf8x]{inputenc}

\title{ Deformed
Super-Yang-Mills in Batalin-Vilkovisky
  Formalism}
\author{Mir Faizal \\ Mathematical Institute
\\ University of Oxford\\ 
Oxford
OX1 3LB}

\begin{document}

\maketitle

\begin{abstract}
 In this paper  we will analyse a three 
dimensional super-Yang-Mills theory on 
 a deformed superspace with boundaries. We show that it is
possible to obtain an undeformed theory on the boundary
 if the bulk superspace is deformed 
by imposing a non-vanishing  commutator
between bosonic  and fermionic coordinates. 
 We will also analyse  this theory  in the Batalin-Vilkovisky   
 (BV) formalism and show that these results also hold at 
a quantum level.
\end{abstract}


\section{Introduction}
The supersymmetric transformation of $\mathcal{N} =1$ 
super-Yang-Mills theory in three dimensions is a total
 derivative, and thus for manifolds with boundaries it 
generates a boundary term \cite{new1}-\cite{new3}. 
 This effectively breaks the supersymmetry of the theory. However, this
 problem can be resolved by imposing boundary conditions.
 Alternately, the action can be made supersymmetric even 
before imposing  boundary conditions. This is
done by the addition of a suitable boundary
 term which exactly 
cancels the  term generated by the
 supersymmetric transformation of the bulk Lagrangian.
Supersymmetry can also be broken by imposing  anti-commutativity
between the Grassman coordinates of the superspace. Such deformation of superspace occur due to 
$RR$ fields  \cite{rr}-\cite{rr1}.   
The deformations of
 a $\mathcal{N}=4$ super-Yang-Mills theory can also occur due  to a 
graviphoton backgrounds  \cite{new4}-\cite{new5}. 
Supersymmetric Yang-Mills theory with chiral 
matter multiplets in presence of graviphoton
 background has also been analysed \cite{new6}. 
In fact, a perturbation theory scheme has been used for  computing correlation
 functions  for this theory. 

As the Lagrangian  for the super-Yang-Mills theory is invariant under 
gauge transformations, so we have to add a gauge fixing term and a ghost term before
doing any calculations. The new Lagrangian thus obtained is invariant under a new set of symmetries 
called the BRST and the anti-BRST symmetries. 
The BRST and the anti-BRST symmetries
 for the ordinary Yang-Mills theory has already been investigated
\cite{16}-\cite{17}. The BRST symmetry of a $\mathcal{N} = 1$ 
   super-Yang-Mills theory 
 has also been studied in  superspace formalism \cite{18}-\cite{19}. 
This theory was analysed in the background field method.
In this  method  all the fields in a theory  are shifted. 
 The BRST 
and the anti-BRST symmetries of these shifted fields is then 
 analysed by using 
Batalin-Vilkovisky   formalism \cite{10a}-\cite{11a}.  
In this formalism the extended BRST and the extended anti-BRST 
symmetries arise  due to the
 invariance of a
 theory under both the original BRST and the original
 anti-BRST transformations along with 
the shift transformation. This  
    have been done for the conventional 
Yang-Mills theories \cite{12a}-\cite{13a}. 
Furthermore, the BRST and the anti-BRST symmetries can be viewed as 
 supersymmetric transformations because
they mix the fermionic and bosonic 
coordinates. 
Thus, the extended BRST and the extended anti-BRST  symmetries 
for Yang-Mills theory have been studied 
  in the extended 
superspace formalism \cite{14a}-\cite{30}. 

In this paper we will analyse the super-Yang-Mills theory 
in presence of boundary. We will show that even if we deform 
the superspace of this original theory, it is possible to obtain a
undeformed theory on the boundary. This is because the boundary only 
preserves half the supersymmetry of the bulk theory. 
Hence, if we perform a deformation in  
the other half, it will not be visible on 
the boundary. We will also analyse
this theory in the BV-formalism and hence show that this 
result holds even at a quantum level.
    
\section{  Deformed Super-Yang-Mills Theory  }
In this section we shall first review the properties 
of the super-covariant derivatives for 
 non-Abelian $\mathcal{N} = 1$ gauge fields in three
 dimensions. In order to
 do that we first introduce  a
two component anti-commuting Grassmann parameters $\theta_a$. We also introduce 
 two anti-symmetric tensors $C^{ab}$ and $C_{ab}$  
which are used to raise and lower spinor indices. These anti-symmetric tensors
satisfy  $C_{ab}C^{bc} = \delta^c_a$. Furthermore, we let 
$
 \theta^2 =  \theta_a C^{ab} \theta_b/2 = 
\theta^a \theta_a/2.
$
Now $\Gamma_a$ is defined to be a matrix valued spinor
 superfield suitable contracted with generators of this Lie algebra, 
 $ \Gamma_a = \Gamma_a^A T_A $, where $T_A$ are Hermitian generators of a Lie algebra 
$[T_A, T_B] = i f_{AB}^C T_C$.  
We also define super-covariant derivatives as 
\begin{equation}
 \nabla_a  = D_a - i \Gamma_a,
\end{equation}
where $D_a$ is the super-derivative  given by 
\begin{equation}
 D_a = \partial_a + (\gamma^\mu \partial_\mu)^b_a \theta_b.
\end{equation}
The matrix valued spinor superfield  transforms under 
gauge transformations as 
\begin{equation}
 \delta \Gamma_a = \nabla_a \Lambda.
\end{equation}
The components of the superfield  $\Gamma_a$ are given by
\begin{eqnarray}
 \chi_a = [\Gamma_a]_|, && A = - \frac{1}{2}[\nabla^a \Gamma_a]_|, 
\nonumber \\ 
A^\mu = - \frac{1 }{2} [ \nabla^a (\gamma^{\mu })_a^b \Gamma_b ]_|, &&
 E_a = - [\nabla^b \nabla_a  \Gamma_b]_|. \label{csf}
\end{eqnarray}
Now let $\mathcal{D}_\mu$ be the conventional covariant derivative given by 
\begin{equation}
 \mathcal{D}_\mu = \partial_\mu -i A_\mu,
\end{equation}
then it can be shown by direct computation 
that the super-covariant derivative  satisfies 
\begin{equation}
 \{\nabla_a , \nabla_b\} = 2 \gamma_{ab}^\mu \mathcal{D}_\mu.
\end{equation}
For a $\mathcal{N} =1$ superfields 
 in three dimensions $[\nabla_a , \nabla_b]$ must be proportional
 to the anti-symmetric tensor $C_{ab}$ \cite{2d}-\cite{2d1}. Thus, we can write 
\begin{eqnarray}
  \nabla_a \nabla_b &=&  \frac{1}{2} \{ \nabla_a, \nabla_b\} + \frac{1}{2}
 [\nabla_a, \nabla_b]  \nonumber \\ &=&  \gamma^\mu _{ab}
 \mathcal{D}_\mu - C_{ab} \nabla^2.
\end{eqnarray}
Furthermore, using \cite{new1}, 
\begin{equation}
D_a D_b D_c = \frac{1}{2} 
D_a \{D_b, D_c\} -  D_b \{D_a, 
D_c \} + \frac{1}{2} D_c \{D_a ,D_b \},
\end{equation}
we find that the super-covariant derivative  satisfies 
\begin{equation}
 \nabla_a \nabla_b \nabla_c = \frac{1}{2} 
\nabla_a \{\nabla_b, \nabla_c\} -  \nabla_b \{\nabla_a, 
\nabla_c \} + \frac{1}{2} \nabla_c \{\nabla_a ,\nabla_b \}. 
\end{equation}
Thus, we get 
\begin{eqnarray}
\nabla^a \nabla_b \nabla_a &=& 0, \label{ar} \label{nab}\\ 
\nabla^2 \nabla_a &=&   (\gamma^\mu \nabla )_{a} \mathcal{D}_\mu,
 \label{q} 
\end{eqnarray}
Now the non-abelian    Yang-Mills   
theory  on this  superspace  can now be
 written as \cite{superym},
\begin{equation}
 \mathcal{L}_{bulk} =\nabla^2 
 Tr (  \omega^a   \omega_a) _|
\end{equation}
where
\begin{eqnarray}
 \omega_a & = & \frac{1}{2} D^b D_a \Gamma_b 
- \frac{i}{2}  [\Gamma^b , D_b \Gamma_a] -
 \frac{1}{6} [ \Gamma^b ,
\{ \Gamma_b , \Gamma_a\} ]. 
\end{eqnarray}
and $'|'$  means that the quantity is evaluated at $\theta_a =0$.
This Lagrangian in component form is given by 
\begin{equation}
  \mathcal{L}_{bulk} = 4 F^{\mu\nu} F_{\mu\nu} +
 2 \overline{\lambda} \gamma^\mu 
\mathcal{D}_{\mu}\lambda. 
\end{equation}
It may be noted that  a theory with $\mathcal{N} =1$ supersymmetry in four dimensions has the same amount of
supersymmetry as a theory with $\mathcal{N} =2$ supersymmetry in three dimensions. 
In four dimensions  two independent super-charges exist, so there are two terms like this.
Similarly, a Yang-Mills   
theory with   $\mathcal{N} =2$ supersymmetry in three dimensions will have two terms like this.
This Lagrangian is invariant under supersymmetric 
transformations generated 
by the supercharge $Q_a$ given by 
\begin{equation}
 Q_a = \partial_a - (\gamma^m \partial_m)^b_a \theta_b.
\end{equation}
It satisfies the following algebra, 
\begin{equation}
 \{ Q_{ a}, Q_{ b} \}=  2  ( \gamma^\mu)_{ab} \partial_\mu.
\end{equation}
 However, in presence of a boundary 
half of this supersymmetry is broken. To understand how this works let us
 put a boundary at fixed $x^3$. 
Now $\mu$ splits into $\mu =(m, 3 )$,
here $m$ labels the coordinates along the 
boundary.
The induced values of the super-derivative $D_a$ and the 
super-covariant derivative $\nabla_a$ on the boundary are denoted by 
$D_a'$ and $\nabla_a'$, respectively.  
This boundary super-derivative $D_a'$ is obtained by neglecting 
$\gamma^3 \partial_3$ contributions in $D_a$, 
 \begin{equation}
 D_a' = \partial_a + (\gamma^m \partial_m)^b_a \theta_b.
\end{equation}
The boundary super-covariant derivative $\nabla_a'$ 
can thus be written using  $\Gamma_a'$  which is the induced values of 
the bulk field $\Gamma_a$ on the 
boundary. Any boundary field along with the 
induced value of any quantity e.g.,  
$\Lambda$ on the boundary will be denoted by $\Lambda'$.
This  convention will be followed even
 for component fields of superfields.  
The  matrix valued spinor superfield 
$\Gamma_a'$ transforms under gauge transformations as
\begin{equation}
 \delta \Gamma'_a = \nabla'_a \Lambda'.
\end{equation}
Now we define a  projection operator $P_{\pm}$ as
\begin{equation}
 (P_{\pm  })^b_a  = \frac{1}{2} (\delta_a^b \pm (\gamma^{3}) ^b_a). 
\label{a}
\end{equation}
The projected values of the 
super-covariant derivative satisfy \cite{2d}-\cite{2d1}
\begin{eqnarray}
 \nabla_{+a} \nabla_{+b} &=& - ( \gamma^+)_{ab} \mathcal{D}_+, \\
 \nabla_{-a} \nabla_{-b} &=& - ( \gamma^-)_{ab} \mathcal{D}_-, \\
 \nabla_{-a} \nabla_{+b} &=&  -(P_-)_{ab} (\mathcal{D}_3 + \nabla^2),\\  
 \nabla_{+a} \nabla_{-b} &=&  (P_+)_{ab} (\mathcal{D}_3 -\nabla^2 ). \label{2}
 \end{eqnarray}
Furthermore,   we can obtain the following 
algebra from these relations,
\begin{eqnarray}
 \{ \nabla_{ + a}, \nabla_{ +b} \} &=& - 2  ( \gamma^+)_{ab} \mathcal{D}_+, \\
  \{ \nabla_{-a }, \nabla_{-b} \} &=& - 2  ( \gamma^-)_{ab} \mathcal{D}_-, \\
 \{ \nabla_{- a}, \nabla_{ +b} \} &=&  -2 
 (P_-)_{ab} \mathcal{D}_3. \label{algebra}
\end{eqnarray}
The generators of the supersymmetry also satisfy a similar 
algebra 
\begin{eqnarray}
 \{ Q_{ + a}, Q_{ +b} \} &=&  2  ( \gamma^+)_{ab} \partial_+, \\
  \{Q_{-a }, Q_{-b} \} &=& 2  ( \gamma^-)_{ab} \partial_-, \\
 \{ Q_{- a}, Q_{ +b} \} &=& 2  (P_-)_{ab} \partial_3. 
\end{eqnarray}

It is possible to add a boundary term to the super-Yang-Mills theory 
to make it preserve half the supersymmetry, even in presence of a boundary
 \cite{new1}. 
This is because we can write the generators of the supersymmetry as 
\begin{equation}
\epsilon^a Q_a = \epsilon^+ Q_+ + \epsilon^- Q_-. 
\end{equation}
Now it is possible to break the supersymmetry corresponding to $Q_-$ 
and retain the supersymmetry corresponding to $Q_+$. This is done 
by addition the following term to the bulk Lagrangian 
\begin{equation}
\mathcal{L}_{boud}= \mathcal{D}_3  Tr (  \omega^a    \omega_a) _|.
\end{equation}
 This term exactly cancels the boundary term generated 
by the supersymmetric variation of the bulk Lagrangian. 
Now we have
\begin{eqnarray}
 - \nabla_+ \nabla_- &=&  -C^{ab}\nabla_{+a} \nabla_{-b}  \nonumber \\
 &=& - C^{ab}(P_+)_{ab} (\mathcal{D}_3 -\nabla^2 ) \nonumber \\
 & = & -(P_+)^a_a(\mathcal{D}_3 -\nabla^2) \nonumber \\ &=&  
(\mathcal{D}_3 -\nabla^2).
\end{eqnarray}
So, we can write the Lagrangian for  supersymmetric  Yang-Mills   
theory  in presence of a boundary as   
\begin{eqnarray}
 \mathcal{L}_{bb} &=& \mathcal{L}_{bulk} + \mathcal{L}_{boud}
\nonumber \\ &=& (\mathcal{D}_3 -\nabla^2) 
Tr (  \omega^a   \omega_a) _|
\nonumber \\ &=& \nabla_+ \nabla_- 
 Tr (  \omega^a  \omega_a) _|.
\end{eqnarray}
This Lagrangian in component form is given by 
\begin{equation}
  \mathcal{L}_c = 4 F^{\mu\nu} F_{\mu\nu} +
 2 \overline{\lambda} \gamma^\mu 
\mathcal{D}_{\mu}\lambda +  \mathcal{D}_3 (\overline{\lambda}\lambda).
\end{equation}

Now we shall deform  this three 
dimensional   Yang-Mills   theory. 
To do so we first let 
$ y^\mu = x^\mu + \theta^a (\gamma^\mu) _{a}^b \theta_b$.  
Then we promote the superspace coordinates 
to operators $\hat{ \theta}^a$ and $\hat{y}^\mu$
 and impose the following 
 deformation of the superspace algebra on them, 
\begin{eqnarray}
 {[\hat{y}^\mu, P_-\hat{\theta}^a]} &=& A^{\mu a }. 
\end{eqnarray} 
This deformation occurs due to a graviphoton  background \cite{new4}-\cite{new6}.
Unlike a $RR$ deformation, this deformation does not break any supersymmetry of the 
theory. 
A matrix valued superfields $\hat{\Gamma}_a 
(\hat{y}, \hat{\theta}) $ on this
 deformed superspace is obtained by suitably contracting 
 it  with generators of the Lie algebra 
$[T_A, T_B] = i f_{AB}^C T_C,$
in the adjoint representation, 
\begin{equation}
 \hat{\Gamma}_a (\hat{y}, \hat{\theta})= 
\hat{\Gamma}_a^A (\hat{y}, \hat{\theta}) T_A.
\end{equation}
We now use Weyl
ordering and express the Fourier transformation of $\hat{\Gamma}_a (\hat{y}, \hat{\theta})$ as \cite{fai1}-\cite{fai2}, 
\begin{equation}
\hat{\Gamma}_a (\hat{y}, \hat{\theta}) =
\int d^3 k \int d^2 \pi e^{-i k \hat{y} -\pi \hat{\theta} } \;
\Gamma_a (k,\pi).
\end{equation}
Now we can also express Fourier transformation of ${\Gamma}_a ({y}, {\theta})$ as, 
\begin{equation}
\Gamma_a (y, \theta)  =
\int d^3 k \int d^2 \pi e^{-i k y -\pi \theta } \;
\Gamma_a (k,\pi).
\end{equation}
Thus,  we  have a one to one map between a function of
the deformed superspace  and a function of ordinary
 superspace. 
 Now we can write  the product of two deformed superfields as
\begin{eqnarray}
{\hat{\Gamma}^a}(\hat{y},\hat{\theta}) { \hat{\Gamma}_{a  }}  
(\hat{y},\hat{\theta}) &=&
\int d^3 k_1 d^3 k_2 \int d^2 \pi_1  d^2 \pi_2\nonumber \\  
&&\times 
\exp -i( ( k_1 +k_2) \hat{y} +(\pi_1 +\pi_2) \hat{\theta} )  \nonumber \\  
&&\times 
 \exp -\frac{i}{2} \left( A^{\mu a} 
(\pi^2 _{-a} k^1_\mu - k^2_\mu \pi^1_{-a} ) \right)  \nonumber \\  
&&\times 
{\Gamma}^a  (k_1,\pi_1) {\Gamma}_{a  }   (k_2,\pi_2).
\end{eqnarray}
This motivates the  definition of the star product
  for the  functions on ordinary
 superspace,  
\begin{eqnarray}
{\Gamma^a}(y,\theta) \star { \Gamma_{a  }}  (y,\theta) & =& 
\exp -\frac{i}{2} \left(
+ A^{\mu a} (\partial^2 _{-a} \partial^1_\mu -
 \partial^2_\mu \partial^1_{-a}) \right) \nonumber \\ && \times
 {\Gamma^a}(y_1,\theta_1) { \Gamma_{a  }}  (y_2, \theta_2)
\left. \right|_{y_1=y_2=y, \; \theta_1=\theta_2=\theta}.
\label{star2}
\end{eqnarray}
 It is also useful to define the following bracket 
\begin{equation}
 [\Gamma^a \star \Gamma_a] = T_A f^A_{BC} \Gamma^{a B} \star \Gamma_a^C.
\end{equation}
In order to construct a  Yang-Mills  
 theory on this deformed superspace,
 it is useful to define 
\begin{eqnarray}
 \omega_a & = & \frac{1}{2} D^b D_a \Gamma_b 
- \frac{i}{2}  [\Gamma^b \star D_b \Gamma_a] -
 \frac{1}{6} [ \Gamma^b \star
[ \Gamma_b \star \Gamma_a] ]. 
\end{eqnarray}
The bulk non-abelian    Yang-Mills   
theory  on this deformed superspace is given by
\begin{equation}
 \mathcal{L}_c =\nabla^2 
 Tr (  \omega^a  \star  \omega_a) _|. 
\end{equation}
So, the boundary theory can be written as 
\begin{equation}
 \mathcal{L}_c =\nabla_+ 
 Tr (  \Omega (\theta_+)) _{\theta_+ =0},
\end{equation}
where
\begin{equation}
 \Omega = \nabla_-(\omega^a  \star  \omega_a)_{\theta_- =0}.
\end{equation}
Thus, we have a undeformed theory on the boundary.
This is because the boundary breaks half of the supersymmetry and we have 
chosen to break the supersymmetry corresponding to $Q_-$, which
is deformed in the bulk.

\section{Batalin-Vilkovisky    Formalism}
Some of the degrees of freedom in the bulk super-Yang-Mills
theory are not physical because  
it is invariant under the following  super-gauge transformations 
\begin{equation}
 \delta \Gamma_a = \nabla_a\star \Lambda, 
\end{equation}
where $\Lambda = \Lambda^A T_A $. 
 This gauge freedom  of the
bulk theory also
induces a gauge freedom for  the boundary theory. So, 
 in order to quantise the boundary
 theory we will have to fix a gauge. This can be done at a quantum level by adding 
 a gauge fixing
 term and a ghost term to the original classical Lagrangian density.  
In order to do that we first define  a matrix  valued scalar superfield
$B$ along with two matrix
 valued anti-commutating superfields $c$ and $\overline{c}$. These superfields are 
suitably contracted with generators of the Lie algebra in the adjoint 
representation, 
\begin{eqnarray}
 F (y, \theta) = F^A(y, \theta) T_A, &&
c(y, \theta) =c^A(y, \theta) T_A, \nonumber \\ 
\overline{c}(y, \theta) = \overline{c}^A(y, \theta) T_A.&&
\end{eqnarray}
Now we can write  the gauge fixing term $\mathcal{L}_{gf}$ and the ghost 
term $\mathcal{L}_{gh}$ as follows:
\begin{eqnarray}
\mathcal{L}_{gf} &=& \nabla_+ Tr 
(\mathcal{N} (\theta_+) )_{\theta_+ =0},
 \nonumber \\
\mathcal{L}_{gh} &=&  \nabla_+  
Tr (\mathcal{K} (\theta_+))_{\theta_+ =0}, 
\end{eqnarray}
where 
\begin{eqnarray}
\mathcal{N} &=& \nabla_- (F  \star D^a \Gamma_a )_{\theta_- =0},
 \nonumber \\
\mathcal{K} &=&  \nabla_-  (\overline{c}  D^a  
\nabla_a  \star c)_{\theta_- =0}.
\end{eqnarray}
The total Lagrangian density which is given by the 
the sum of the original Lagrangian density with 
the gauge fixing term and the ghost term
\begin{equation}
 \mathcal{L} = \mathcal{L}_c  +  \mathcal{L}_{gf} + \mathcal{L}_{gh}.
\label{tl}
\end{equation}
The  BRST transformations are given by 
\begin{eqnarray}
s \Gamma_a=   \nabla_a \star  c, &&
s c = -\frac{1}{2} [c\star c],     \\ \nonumber
s \overline{c}=-   F, &&
s F=0,
\end{eqnarray}
and the anti-BRST transformations are given by 
\begin{eqnarray}
\overline s \Gamma_a =   (D_a -i \Gamma_a)   \star \overline{c}, &&
\overline s c = F- [c \star \overline c], \\ \nonumber
\overline s \overline c = -\frac{1}{2}   [ \overline{c}\star \overline{c}], &&
\overline s F= [F \star \overline c],
\end{eqnarray}
where 
\begin{equation}
 [c\star c] =  T_A f^A_{BC}  c^B \star c^C.
\end{equation}
Both these transformations are nilpotent, 
$
 s^2 = \overline s ^2 =0.
$
In fact, they also satisfy $s \overline s + \overline s s =0$. 
The sum of the gauge fixing term
  and the ghost term  can be expressed as 
\begin{eqnarray}
\mathcal{L}_{gf} + \mathcal{L}_{gh} & =& - \nabla_+   
  \frac{s \overline s }{2} Tr ( \mathcal{Z} (\theta_+))_{\theta_+ =0}
 \nonumber \\ 
 & =& \nabla_+  \frac{\overline{s} s }{2} Tr 
 ( \mathcal{Z}(\theta_+) )_{\theta_+ =0}.
\end{eqnarray}
where
\begin{equation}
 \mathcal{Z} = \nabla_-(\Gamma^a \star \Gamma_a)_{\theta_- =0}.
\end{equation}
So, we can obtain undeformed  gauge fixing and ghost terms 
for the boundary theory, even though the bulk theory is 
 deformed. 
Furthermore, the total 
 Lagrangian density is invariant under the  BRST and the anti-BRST transformations. 
This is because the BRST and the anti-BRST transformations 
are nilpotent and the sum of the gauge fixing 
term and the ghost term can be written as a total
 BRST and  a total anti-BRST variation. Hence, any BRST or  anti-BRST transformation
of the sum of the gauge fixing 
term and the ghost term vanishes. The BRST or the anti-BRST transformations of 
 the original classical Lagrangian density
 are ghost or anti-ghost valued gauge transformations and so they also vanish. Hence, 
the total Lagrangian density is invariant under both the BRST and the anti-BRST transformations, 
$ s \mathcal{L} =\overline{s} \mathcal{L} =0$. 
Now we let 
\begin{eqnarray}
\mathcal{X} & =& -  \frac{s \overline s }{2} \mathcal{Z} (\theta_+) 
\nonumber \\ 
 & =&   \frac{\overline{s} s }{2} 
 \mathcal{Z}(\theta_+).  
\end{eqnarray}
Now the total Lagrangian density given by Eq. (\ref{tl}), can be written
as
\begin{equation}
 \mathcal{L}_c =\nabla_+ 
 Tr (  \mathcal{X}(\theta_+) +  \Omega (\theta_+)) _{\theta_+ =0}.
\end{equation}
So, the deformation of the bulk theory does not effect the boundary 
theory even at a quantum level. 

We will analyse this theory in background field method. So, we start by 
 shift all the original fields as
\begin{eqnarray}
 \Gamma_a &\to&  \Gamma_a - \tilde{ \Gamma }_a, \nonumber \\
c &\to& c - \tilde{c}, \nonumber \\
\overline{c} &\to& \overline{c} - \tilde{\overline{c}},\nonumber\\
F &\to&  F - \tilde{ F}.
\end{eqnarray}
Now we require the Lagrangian density to be 
 invariant under both the original
 BRST transformations and these shift
transformations,
\begin{equation}
\tilde{\mathcal{ L}}=\mathcal{L}(\Gamma_a - 
\overline {\tilde \Gamma_a}, c- \overline{\tilde c}, \overline{c}-\overline{\tilde c}, F- \tilde F).
\label{9a}
\end{equation}
It will be useful to define  $\tilde{\nabla}_a$ as 
\begin{equation}
 \tilde{\nabla}_a = D_a -i \Gamma_a + i \tilde{\Gamma}_a. 
\end{equation}
The 
extended BRST  transformations are given by 
\begin{eqnarray}
s \Gamma_a= \psi_{a}, &&
s \tilde \Gamma_a= \psi_{a} -\tilde{\nabla}_a \star (c- \tilde c),\\ \nonumber
s c =   \epsilon &&  s \tilde c=  \epsilon + \frac{1}{2}[(c- \tilde c) \star ( c- \tilde c)],\\ 
\nonumber
s \overline{c} =    \overline{\epsilon},  &&
s \tilde{\overline{c}}= \overline{\epsilon} + (F-\overline{F}), \nonumber \\ 
s F =\psi,  &&
s \tilde F=\psi,
\end{eqnarray}
where 
\begin{equation}
 \tilde{\nabla}_a \star (c- \tilde c) = D_a \star (c- \tilde c)  - i (\Gamma_a - \tilde{\Gamma}_a)
 \star (c- \tilde c), 
\end{equation}
and $\psi_a, \epsilon, \overline{\epsilon}, \psi$ are the ghost superfields associated with the 
shift symmetries of the original fields
 $ \Gamma_a, c,
 \overline{c}, F$, respectively.
The  BRST transformations  of these superfields vanish, 
\begin{eqnarray}
s\psi_{a} =0, && 
s\epsilon =0,\\ \nonumber
s \overline{\epsilon}=0, &&
s \psi=0,
\end{eqnarray}
Now we define anti-fields $\Gamma^{*}_{a}, c^*, \overline{c}^*$ and $ F^*$, such that 
they  have  opposite parity to the original fields.
The BRST transformations of these anti-fields is given by 
\begin{eqnarray}
s \Gamma^{*}_{a}=-   b_a, &&  
s c^{*}=- B,\\ \nonumber
s \overline c^{*}=-   \overline{B}, && 
s F^{*}=-   b,
\end{eqnarray}
where   
$b_a, B,  \overline{B}$ and $b $ are 
 new auxiliary superfields. The BRST transformations of these new auxiliary superfields vanishes
\begin{eqnarray}
s b_a=0, &&
s B=0,\nonumber \\
s \overline{B}=0, &&
s b=0.\label{3b}
\end{eqnarray}
The  Lagrangian density to
 gauge fix the shift symmetry is chosen to make the 
tilde fields  to
 vanish, so as to obtain the original  Lagrangian density,
\begin{eqnarray}
 \tilde {\mathcal{L}}_{gf}+  
\tilde {\mathcal{L}}_{gh} = \nabla_+ 
  Tr(\mathcal{W}(\theta_+) )_{\theta_+ =0},
\end{eqnarray}
where 
\begin{eqnarray}
\mathcal{W} &=&\nabla_-( -b^a    \star \tilde{\Gamma}_{a} + \Gamma^{*a} \star 
(\psi_{a}-\tilde{\nabla}_a  \star (c- \tilde{c}))\nonumber \\ &&
- \overline B \star  \tilde{c}+\overline{c}^{*}\star \left(\epsilon+\frac{1}{2}
[(c-\tilde{c}) \star (c-\tilde{c})]\right)\nonumber \\ &&
+B \star \tilde{\overline{c}}-c^{*} \star (\overline{\epsilon}+(F-\tilde{F})+b 
  \star \tilde{F}\nonumber \\ && 
+F^{*}\star \psi)_{\theta_- =0}.
\end{eqnarray}
The  tilde fields  vanish upon integrating out the auxiliary superfields.
 This Lagrangian density is  invariant under the original
 BRST transformation and the shift
 transformations.
We also have 
the original Lagrangian density, which is only a
 function of the original fields,
\begin{equation}
\mathcal{L}'_{gf} + \mathcal{L}'_{gh} =  \nabla_+   
Tr\left(\mathcal{U} (\theta_+)\right)_{\theta_+ =0}, 
\end{equation}
where 
\begin{equation}
 \mathcal{U} =\nabla_-\left( -\frac{\delta  \Psi}{ \delta \Gamma_a}\star 
\psi_{a}+\frac{\delta  \Psi}{ \delta c}\star \epsilon
 +\frac{\delta  \Psi}{ \delta  \overline{c}}\star \overline{\epsilon}-\frac{\delta \Psi}{\delta F}\star
 \psi\right)_{\theta_- =0}.
\end{equation}
The classical Lagrangian density can now be written as
\begin{equation}
 \mathcal{L}_c =
 \nabla_+   Tr ( \overline \Omega (\theta_+))_{\theta_+ =0}, 
\end{equation}
where 
\begin{equation}
 \overline \Omega 
= \nabla_-((\omega^a - \tilde\omega^a  )\star  (\omega_a-
\tilde\omega_a))_{\theta_- =0}.
\end{equation}
Now the total Lagrangian density is given by
\begin{eqnarray}
 \mathcal{L}&=&\mathcal{L}_c 
 +\tilde {\mathcal{L}}_{gf}+ \tilde{\mathcal{L}}_{gf} 
+ \mathcal{L}'_{gf} 
+ \mathcal{L}'_{gh} 
\nonumber  \\  &=&
\nabla_+ Tr( \overline \Omega (\theta_+) 
+ \mathcal{U} (\theta_+)
+ \mathcal{W} (\theta_+))_{\theta_+ =0}.
\end{eqnarray}
If we integrate out the fields,
 setting the tilde fields to zero, we have
\begin{equation}
\mathcal{L}= 
\nabla_+   Tr ( \mathcal{M}(\theta_+) )_{\theta_+ =0}, \label{1A}
\end{equation}
where 
\begin{eqnarray}
 \mathcal{M}  &=& 
\nabla_- \left( \Omega + \Gamma^{*a}\star \nabla_a  
c+ \frac{1}{2}\overline{c}^{*}\star
 [c  \star c]  -c^{*}\star 
F\right. \nonumber \\ && \left. -\left(\Gamma^{*a}  +\frac{\delta \Psi}{\delta \Gamma_a}\right)\star \psi^{a} 
+ \left(\overline{c}^{*} + \frac{\delta  \Psi}{ \delta c}\right)\star\epsilon-\left(c^{*} + 
\frac{\delta  \Psi}{ \delta  \overline{c}}\right)
\star\overline{\epsilon} \right. \nonumber \\ && \left. +\left(F^{*}- 
\frac{\delta \Psi}{\delta F}\right)\star \psi  
\right)_{\theta_- =0}.
\end{eqnarray}
Thus, even after shifting all the fields  we obtained a undeformed boundary theory,  even though our 
 bulk theory was deformed. This result even holds at 
 a quantum level. By  integrating out the ghosts associated with
 the shift symmetry, we obtain explicit expression for the anti-fields,
\begin{eqnarray}
 \Gamma^{*a}= -\frac{\delta\Psi}{\delta \Gamma_a},&&
\overline{c}^*= -\frac{\delta \Psi}{\delta c},\nonumber\\
c^{*}=\frac{\delta \Psi}{\delta \overline{c}},&&
F^{*}=\frac{\delta \Psi}{\delta F}.
\end{eqnarray}
With these identifications we obtain an explicit form
 for the total Lagrangian density in background field method.

 This Lagrangian density  is invariant under the 
extended BRST transformations. 
However, the  original Lagrangian density was also  
  invariant under the    
 anti-BRST transformations. So, this Lagrangian density, must also be invariant 
under  on-shell extended anti-BRST transformation as the on-shell
 extended anti-BRST transformations  reduce to the original  
anti-BRST transformations \cite{10mfmk}. The extended anti-BRST transformation of 
the original superfields is given by 
\begin{eqnarray}
\overline{s} \Gamma_a&=  &  \Gamma^{*}_{a} +\tilde{\nabla}_a \star (c-\tilde{\overline{c}}), \nonumber \\ 
\overline{s}c&=   & c^{*}+(F-\overline{F})-   [(c -\tilde{c}) \star (\overline{c}-\tilde{\overline{c}})], \nonumber \\ 
\overline{s}\overline{c}&=  &  \overline{c}^{*}- \frac{1}{2}[(\overline{c}-\tilde{\overline{c}}) \star
 ( \overline{c}-\tilde{\overline{c}})], 
\nonumber \\ 
\overline{s} F &= &   F^*  
  + [(F - \tilde{F}) \star ( \overline{c}-\tilde{\overline{c}})], 
\end{eqnarray}
and the extended anti-BRST transformations of the shifted superfields
is given by
\begin{eqnarray}
 \overline{s}\tilde{\Gamma_a}=  \Gamma^{*}_{a}, &&
 \overline{s}\tilde{c}=    c^{*},\\ \nonumber
\overline{s}\tilde{\overline{c}}=     \overline{c^{*}}, &&
 \overline{s}\tilde{F}=    F^{*}.
\end{eqnarray}
The anti-BRST transformations of the ghost superfields is given by,
\begin{eqnarray}
\overline{s} \psi_a & =& b_a + \tilde{\nabla}_a  \star( F -\tilde{F} ) - [(\tilde{\nabla}_a \star ( c -\tilde c)) \star  
( \overline{c}-\tilde{\overline{c}})],\\ \nonumber
\overline{s} \epsilon&=&   B - [(F-\tilde{F})\star( \overline{c}-\tilde{\overline{c}})]
+[[(\overline{c}-\tilde{\overline{c}})\star(c-\tilde{c})] \star (c-\tilde{c})], \\ \nonumber
\overline{s}\overline{\epsilon}&=&    \overline{B} - [(F-\tilde{F})\star( \overline{c}-\tilde{\overline{c}})],\\ \nonumber
\overline{s} \psi&=&   b,
\end{eqnarray}
The extended anti-BRST transformations of the anti-fields and the new
auxiliary fields  vanishes, 
\begin{eqnarray}
\overline{s }b_a =0, && \overline{s}  \Gamma ^{*}_a =0,\nonumber\\
\overline{s} B=0, &&\overline{s} c^{*}=0,\nonumber\\
\overline{s} \overline{B}=0, &&\overline{s} \overline c^{*}=0,\nonumber\\
\overline{s }\overline{b}=0, && \overline{s} F^{*}=0.
\end{eqnarray}
So, we obtained an undeformed   classical 
Lagrangian density,  gauge fixing 
term and the ghost term for the boundary theory, even though the bulk theory was deformed. 
Furthermore,  these results also hold at 
a quantum level.

\subsection{Conclusion}
In this paper we analysed the 
Yang-Mills theory in a deformed superspace, where 
the deformation was caused by  noncommutativity
between bosonic  and fermionic coordinates. 
We deformed the bulk theory in 
such a way that we could obtain an undeformed theory on the boundary.
We analysed  the  BRST
and the anti-BRST  symmetries of this boundary theory in the
BV-formalism. The ghost and gauge fixing 
terms for the boundary theory   were not 
effected by deformations in the bulk. 
This implied that this result also hold at a quantum level.

The spacelike noncommutative field 
theories are known to be unitarity \cite{u}-\cite{u2}.  
However,  due to Eq. $(\ref{star2})$,
 higher order temporal derivatives will occur in the product of fields 
due to  spacetime noncommutativity. 
It is well known that
the evolution of the $S$-matrix is not unitary  for the 
field theories with higher order temporal derivatives
 \cite{3an}-\cite{14n}.
Thus, spacetime noncommutativity  will  
 break the unitarity of the resultant theory. 
However, if we restrict the theory to spacelike
 noncommutativity and thus do not include any 
higher order temporal derivatives, then this problem can be avoided. 
In fact, in the case of spacelike noncommutativity it is possible 
to construct the Norther's charges \cite{nm1}-\cite{nm2}.
Thus, if we restrict the  spacetime 
deformations to spacelike noncommutativity, then 
 we can construct the Norther's charges for this deformed theory.
 It will be interesting to construct the BRST and the 
anti-BRST charges for this theory 
and use them to find the physical states of this theory.

\end{document}